\documentclass[a4paper,11pt]{article}
\usepackage[utf8]{inputenc}

\usepackage{graphicx}
\usepackage[paper=a4paper,left=25mm,right=25mm,top=22mm,bottom=22mm]{geometry}
\usepackage{amsthm}
\usepackage{amssymb}
\usepackage{amsmath}
\usepackage{hyperref}
\usepackage{xcolor}
\usepackage{enumerate}
\usepackage{listings}
\usepackage{complexity}
\usepackage{microtype}
\usepackage{todonotes}
\usepackage[inline]{enumitem}

\usepackage[font=small,labelfont=bf]{caption}
\usepackage[font=small,labelfont=normalfont,labelformat=simple]{subcaption}

\graphicspath{{./figures/}}

\newtheorem{theorem}{Theorem}

\newtheorem{conjecture}{Conjecture}

\DeclareMathOperator{\conv}{conv}

\def\inst#1{$^{#1}$}

\date{}

\title{Blocking Delaunay Triangulations from the Exterior\footnote{
	We thank anonymous reviewers for valuable comments.
	Parada was supported by the Austrian Science Fund (FWF): W1230 and by Independent Research Fund Denmark grant 2020-2023
	(9131-00044B) ``Dynamic Network Analysis''.
	Scheucher, Parada, and Vogtenhuber were partially supported within the collaborative D-A-CH project \emph{Arrangements and Drawings}, by grants DFG: FE~340/12-1 and FWF: I~3340-N35, respectively.
	Scheucher was supported by the DFG Grant SCHE~2214/1-1.
}}

\begin{document}
	
	\author{
		Oswin Aichholzer\inst{1}
		\and
		Thomas Hackl\inst{1}
		\and
		Maarten L\"offler\inst{2}
		\and
		Alexander Pilz\inst{1}
		\and
		Irene Parada\inst{3}
		\and
		Manfred Scheucher\inst{4}
		\and
		Birgit Vogtenhuber\inst{1}
	}

	\maketitle

	\begin{center}
		{\footnotesize
			\inst{1} 
			Institute of Software Technology, Graz University of Technology, Austria \\
			\texttt{oaich@ist.tugraz.at}, \texttt{bvogt@ist.tugraz.at}
			\\\ \\
			\inst{2} 
			Department of Information and Computing Sciences,\\
			Utrecht University, Netherlands,\\
			\texttt{m.loffler@uu.nl}
			\\\ \\
			\inst{3} 
			Department of Applied Mathematics and Computer Science,\\
			Technical University of Denmark\\
			\texttt{irmde@dtu.dk}
			\\\ \\
			\inst{4} 
			Institut f\"ur Mathematik, Technische Universit\"at Berlin, Germany\\
			\texttt{lastname@math.tu-berlin.de}
			\\\ \\
		}
	\end{center}

\begin{abstract}
	Given two distinct point sets $P$ and $Q$ in the plane, 
	we say that $Q$ \emph{blocks} $P$ if no two points of~$P$ are adjacent in any Delaunay triangulation of $P\cup Q$.
	Aichholzer et al.\ (2013) showed that
	any set $P$ of $n$ points in general position
	can be blocked by $\frac{3}{2}n$ points and that every set $P$ of $n$ points in convex position can be blocked by $\frac{5}{4}n$ points.
	Moreover, they conjectured that, if $P$ is in convex position, $n$ blocking points are sufficient and necessary.
	The necessity was recently shown by Biniaz (2021)
	who proved that every point set in general position 
	requires $n$ blocking points.
	
	Here we investigate the variant, where blocking points can only lie outside of the convex hull of the given point set. 
	We show that $\frac{5}{4}n-O(1)$ such \emph{exterior-blocking} points are sometimes necessary, even if the given point set is in convex position.
	As a consequence we obtain 
	that, if the conjecture of Aichholzer et al.\ 
	for the original setting was true,
	then minimal blocking sets of some point configurations~$P$ would have to contain points inside of the convex hull of~$P$.
\end{abstract}


\section{Introduction}

\begin{figure}[htb]
	
	\hbox{}\hfill
	\begin{subfigure}[t]{.3\textwidth}
		\centering
		\includegraphics[page=1]{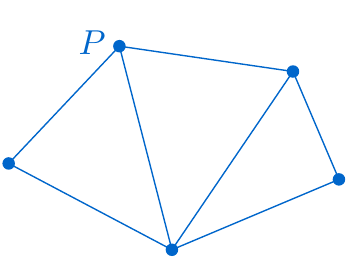}
		\caption{}
		\label{fig:example_a}  
	\end{subfigure}
	\hfill
	\begin{subfigure}[t]{.3\textwidth}
		\centering
		\includegraphics[page=2]{figs/example}
		\caption{}
		\label{fig:example_b}  
	\end{subfigure}
	\hfill
	\begin{subfigure}[t]{.3\textwidth}
		\centering
		\includegraphics[page=3]{figs/example}
		\caption{}
		\label{fig:example_c}  
	\end{subfigure}
	\hfill\hbox{}
	
	\caption {\subref{fig:example_a}~A set $P$ (blue) of five points in convex position, and its unique Delaunay triangulation $DT(P)$. \subref{fig:example_b}~A set $Q$ (red) of two points that blocks two of the edges of $DT(P)$. 
		\subref{fig:example_c}~A~set $Q$ of five points from the exterior of $\conv(P)$ that blocks~$P$.}
	\label {fig:example}
	\end {figure}

Delaunay triangulations, Delaunay graphs, Voronoi diagrams (their dual structures), and various generalizations have been intensively studied in the last century;
see for example the standard textbook in Computation Geometry \cite{CGbook}.
A \emph{Delaunay triangulation} $DT(P)$ of a given point set $P$ in the plane 
is a triangulation of $P$ in which for
every edge between two distinct points $p_1,p_2 \in P$ there exists a circle through $p_1,p_2$ that contains no point of $P\setminus\{p_1,p_2\}$ in its interior. 
An edge spanned by $P$ with this property is called \emph{Delaunay edge}.
For a point set \emph{in general position}, that is, 
no three points of $P$ lie on a common line and no four points of $P$ lie on a common circle, the Delaunay triangulation is unique. 
Figure~\ref{fig:example_a} shows the unique Delaunay triangulation of 
a point set \emph{in convex position}, 
that is, the points are the vertices of a convex polygon.

	In this article we continue the investigation of blocking points for Delaunay edges.
	For two point sets $P,Q$,
	we say that $Q$ \emph{blocks an edge} $p_1p_2$ spanned by $P$ if  
	every circle through $p_1$, $p_2$ 
	contains at least one point of $P \cup Q$ in its interior.
	Equivalently, $p_1p_2$ is not an edge of any Delaunay triangulation of $P\cup Q$.
	We say that $Q$ \emph{blocks $P$} if $Q$ blocks all edges spanned by~$P$. 
	Equivalently, no two points of $P$ are adjacent in any Delaunay triangulation of $P\cup Q$. 
	If moreover no point of $Q$ lies in the interior of the convex hull of $P$, 
	we say that $Q$ \emph{blocks $P$ from the exterior}.
	Figures~\ref{fig:example_b} and~\ref{fig:example_c} shows examples where $Q$ blocks (parts of) $P$.

	Aronov et al.\ \cite{ADH2011} showed that 
	every set $P$ of $n$ points in general position 
	can be blocked by a set of $2n-2$ points,
	and that, if $P$ is in convex position, $\frac{4}{3}n$ blocking points are sufficient.
	Both of their bounds were improved by
	Aichholzer et al.\ \cite{AFMHKPRV2013}, who showed that, 
	for general position, $\frac{3}{2}n$ blocking points are sufficient,
	and that, for convex position, $\frac{5}{4}n$ blocking points are sufficient.
	They also showed that $n-1$ blocking points are always needed
	and posed the following conjecture.
	\begin{conjecture}[\cite{AFMHKPRV2013}]
		\label{conj:convex_n_blocking}
		If $P$ is a set of $n$ points in convex position in the plane,
		then $n$ blocking points are necessary and sufficient, that is, 
		every blocking set of $P$ contains at least $n$ points and this bound is tight.
	\end{conjecture}
	Biniaz \cite{Biniaz2021} recently 
	strengthened the lower bound by showing that, for every set of $n$ points in general position,
	$n$ blocking points are necessary 
	and that there are sets of $n$ points in convex position which can be blocked by $n$ points.
	While this confirms the necessity part from Conjecture~\ref{conj:convex_n_blocking},
	the question about sufficiency remains open.
		
	For many sets $P$ of $n$ points in convex position, a simple construction suffices to indeed block all Delaunay edges with exactly $n$ points: place a single point of $Q$ close to the mid point of each edge of the convex hull of $P$, on the outer side; see Figure~\ref{fig:example_c}. Placing the points 
	arbitrary close to the convex hull edges
	ensures that all those edges are blocked, and indeed every convex hull edge requires at least one point somewhere outside the convex hull to be blocked. 
	Moreover, this simple construction often enough also blocks all interior edges of $DT(P)$. This 
	may suggest that a similar approach could actually always work.
	
	\goodbreak
	
	Inspired by these observations, we investigate the variant 
	where blocking points have to lie outside of the convex hull of the given $n$-point set $P$.
	We show that $\frac{5}{4}n-O(1)$ such \emph{exterior-blocking} 
	points are sometimes necessary, even if $P$ is in convex position.
	
	\goodbreak
	
	\begin{theorem}
		\label{thm:genpos_thm}
		For $k \in \mathbb{N}$,
		there is a set $P$ of $4k$ points in general position
		that requires at least $5k-5$ exterior-blocking points.
	\end{theorem}
	
	As a direct consequence of Theorem~\ref{thm:genpos_thm} 
	we obtain
	for the original setting that,
	if Conjecture~\ref{conj:convex_n_blocking} was true,
	then minimal blocking sets of certain point sets~$P$ 
	would have to contain points inside of the convex hull of~$P$. 
	
	Note that the construction of size $\lfloor \frac{5}{4}n \rfloor$ for convex position in~\cite{AFMHKPRV2013} might contain interior points. The reason is that in the induction blocking points placed for a subproblem in the exterior of an edge (Case (a) in the proof of Theorem 3 in~\cite{AFMHKPRV2013}) might end up to be interior for the overall triangulation. Modifying their approach, a blocking set of size $\approx \frac{4}{3}n$ can be obtained by iteratively cutting ears ($(n,3,4)$-cuts in the terminology of~\cite{AFMHKPRV2013}).

	\section{Proof of Theorem~\ref{thm:genpos_thm}}

	To prove Theorem~\ref{thm:genpos_thm},
	we first give a configuration with collinear points in Section~\ref{sec:collinear},
	which we then perturb in Section~\ref{sec:genpos} 
	to obtain a configuration which is in general position.

	\subsection{Construction with Collinear Points}
	\label{sec:collinear}

\begin{figure}[htb]
		\centering
		\includegraphics[page=3,width=0.95\textwidth]{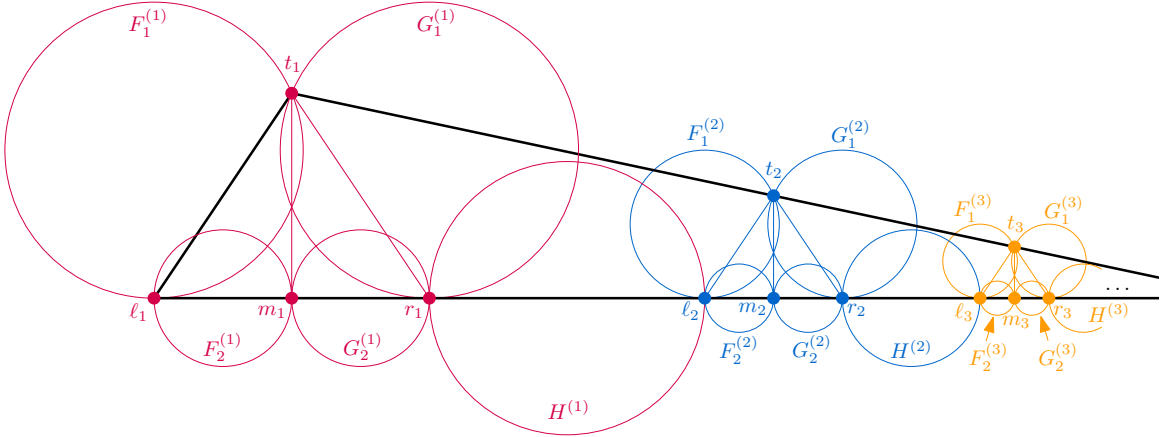}
		\caption{An illustration of the point set $P_0$ of size $4k$ and the set of circles $\mathcal{C}_0$ 
			where at least $5k-3$ exterior-blocking points are required. 
			The red, blue, and yellow points and circles illustrate the first, second, and third gadget of the construction, respectively.}
		\label{fig:collinear_construction}
	\end{figure}

	Our construction consists of $k$ gadgets, each containing $4$ points:
	a \emph{top point} $t_i$, a \emph{left point} $\ell_i$, \emph{middle point} $m_i$, and a \emph{right point} $r_i$, where the latter three are called \emph{bottom points}. This gives us a set $P_0$ of $n=4k$ points in total.
	We place all $3k$ bottom points on the $x$-axis
	and all $k$ top points on a line segment (above the $x$-axis) with negative slope; c.f.~Figure~\ref{fig:collinear_construction}. 
	
	Explicit coordinates for the points $\{\ell_i,m_i,r_i,t_i\}$ in the $i$-th gadget are 
	$\{(-2,0),(0,0),\allowbreak{}(2,0),(0,3)\}$, 
	scaled by $2^{-i}$, and with $x$-offset of $3+14 \sum_{j = 1}^{i} 2^{-j} = 3+14 (1-2^{-i})$. 
	By construction, all points have positive $x$-coordinate,
	all bottom points lie on the $x$-axis,
	and all top points lie on the line 
	$\{
	(x,y) : 3x + 14y = 51 
	\}$.
	
	Further, each gadget $i$ 
	with $1 \le i < k$ contains 5 circles 
	and the $k$-th gadget contains 4 circles, which gives us a set $\mathcal{C}_0$ of $5k-1$ circles in total. 
	They are defined as follows: 
	\begin{itemize}
		\item
		a circle $F_1^{(i)}$ through $t_i$ and~$\ell_i$, 
		which is tangent to the $x$-axis in~$\ell_i$;
		
		\item
		a circle $G_1^{(i)}$ through $t_i$ and~$r_i$, 
		which is tangent to the $x$-axis in~$r_i$,
		
		\item
		a circle $F_2^{(i)}$ with the segment $\ell_i m_i$ as diameter,
		
		\item
		a circle $G_2^{(i)}$ with the segment $m_i r_i$ as diameter; and
		
		\item
		a circle $H^{(i)}$ with the segment $r_i \ell_{i+1}$ as diameter.
	\end{itemize}
	
	See Figure~\ref{fig:collinear_construction} for an illustration of the construction.
	On each circle, there are exactly two points of $P_0$ and no circle contains points of $P_0$ in its interior.
	Further, any two ``neighboring'' bottom circles are tangent in their common point of~$P_0$,
	that is, 
	$F_{2}^{(i)} \cap G_{2}^{(i)} = \{m_i\}$,
	$G_{2}^{(i)} \cap H_{2}^{(i)} = \{r_i\}$, and
	$H_{2}^{(i)} \cap F_{2}^{(i+1)} = \{\ell_{i+1}\}$.

	It is necessary that each of the circles contains a blocking point of $Q$ in its interior
	as otherwise there is an edge in the Delaunay graph of $P_0$ and hence in any Delaunay triangulation.
	For each circle $C$, we denote the region in the interior of $C$ and in the exterior of the convex hull of $P_0$ as its \emph{blocking area}.
	Note that the circles
	$F_{1}^{(i)}$ and $G_{1}^{(i)}$ are both tangent the x-axis and thus only contain points above the x-axis in their interior,
	and that the circles $H^{i}$ can only be blocked from points below the x-axis.
	Therefore no two circles (except in the first and last gadget) 
	have a common exterior-blocking area.
	Therefore, five exterior-blocking points are required to block all circles of a gadget for $1 < i < k$.
	For the first and last gadget, 4 and 3 exterior-blocking points are required, respectively. As none of these points can be used for two gadgets simultaneously, a total of $5k-3$ points is required to block $P_0$ from the exterior.

	\subsection{Transformation to General Position}
	\label{sec:genpos}
	
\begin{figure}[htb]
		
		\begin{subfigure}[t]{.47\textwidth}
			\centering
			\includegraphics[page=1,width=\textwidth]{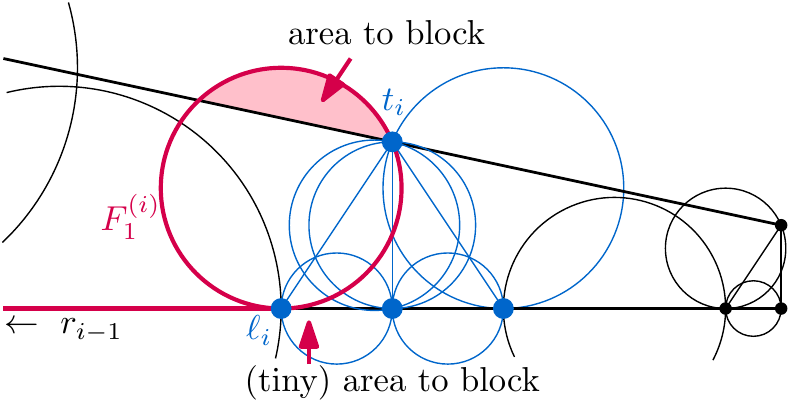}
			\caption{}
			\label{fig:general_construction_circle1}  
		\end{subfigure}
		\hfill
		\begin{subfigure}[t]{.47\textwidth}
			\centering
			\includegraphics[page=2,width=\textwidth]{figs/general_construction}
			\caption{}
			\label{fig:general_construction_circle2}  
		\end{subfigure}
	
		\bigskip
		
		\begin{subfigure}[t]{.47\textwidth}
			\centering
			\includegraphics[page=3,width=\textwidth]{figs/general_construction}
			\caption{}
			\label{fig:general_construction_circle3}  
		\end{subfigure}
		\hfill
		\begin{subfigure}[t]{.47\textwidth}
			\centering
			\includegraphics[page=4,width=\textwidth]{figs/general_construction}
			\caption{}
			\label{fig:general_construction_circle4}  
		\end{subfigure}
		
		\caption{The gadget for the general case construction. 
			\subref{fig:general_construction_circle1} -- \subref{fig:general_construction_circle4} show how to align circles 
			(the red circle is always tangent to the red line)
			and highlight the exterior blocking area using red arrows.
		}
		\label{fig:general_construction}
	\end{figure}

	We will 
	slightly perturb the point set $P_0$ such that all points are in convex position.
	We also add two more circles for each gadget $i$ 
	with $1 < i < k$ to the set $\mathcal{C}_0$ and remove the circles $F_1^{(i)}$ and $G_1^{(i)}$ for $i=1,k$. 
	We denote the resulting set or circles by~$\mathcal{C}_0'$. The new circles are defined as follows; see Figures \ref{fig:general_construction_circle2} and~\ref{fig:general_construction_circle3} for an illustration.
	\begin{itemize}
		\item
		a circle $F_3^{(i)}$ through $t_i$ and~$m_i$, 
		which is tangent to the segment $t_it_{i+1}$; and
		
		\item
		a circle $G_3^{(i)}$ through $t_i$ and~$m_i$,   
		which is tangent to the segment $\ell_i m_i$.
	\end{itemize}

	Note that a circle $C$ through a point $p$ cannot simultaneously be tangent to two line segments at $p$ with different slopes. 
	Thus, the arguments from Section~\ref{sec:collinear} will not apply anymore after we perturb $P_0$, because circles $F_1^{(i)}$ and $G_1^{(i)}$ will intersect other circles outside the convex hull of $P_0$. 
	In the following we will deal with this issue.

\begin{figure}[htb]
		\centering
		\includegraphics[width=0.95\textwidth]{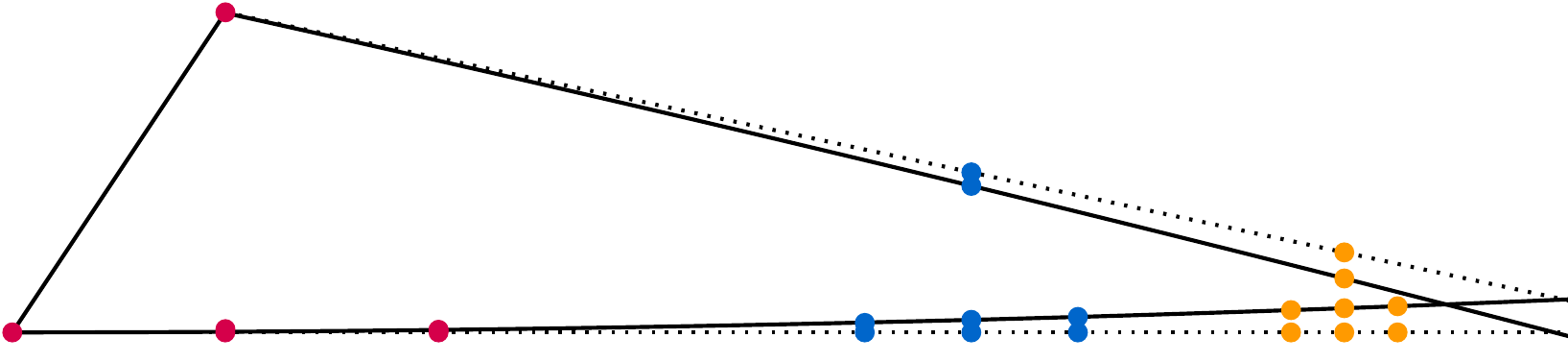}
		\caption{An illustration of the point set $P_0 = P(0)$ and the perturbed set $P(\tau)$ for sufficiently small~$\tau$. Note that, if $\tau$ is not small enough, the resulting set might not be in convex position and hence might not have the desired properties .}
		\label{fig:collinear_construction_perturbation}
	\end{figure}

	\subparagraph*{Transformation.}
	We define $P(\tau)$ as the continuous transformation of $P_0=P(0)$ where 
	\begin{itemize}
		\item 
		all bottom points are transformed as $(x,y) \mapsto (x,y+\tau x^3)$ and 
		\item
		all top points are transformed as $(x,y) \mapsto (x,y-\tau x^3)$.
	\end{itemize}
	The transformation 
	is illustrated in Figure~\ref {fig:collinear_construction_perturbation}.
	
	Analogously, we define $\mathcal{C}(\tau)$ as the transformation of $\mathcal{C}_0'$,
	which preserves the defined properties of the circles, where for $1<i<k$, we keep the tangency of $F_1^{(i)}$ with $r_{i-1}\ell_i$ and the one of $G_1^{(i)}$ with $r_i\ell_{i+1}$. See Figures \ref{fig:general_construction_circle1} and~\ref{fig:general_construction_circle4}.
	Since all circles in $\mathcal{C}_0'$ have finite radii,
	we can choose $\tau_{max} >0$ such that
	all points of $P(\tau)$ are in general position and 
	lie on the boundary of the convex hull 
	and all circles of $\mathcal{C}(\tau)$ have finite radii for $0 \le \tau \le \tau_{max}$.
	Details are deferred to Section~\ref{app:details}.
	
	In the following, we denote by $c(C)$ the center of a circle $C$ and by $r(C)$ the radius of $C$, 
	and we define $d_{C,p} := \| p-c(C) \| - r(C)$ to indicate whether the point $p$ lies 
	\begin{itemize}
		\item 
		inside the circle $C$ ($d_{C,p}<0$), 
		\item 
		on the circle $C$ ($d_{C,p}=0$), or 
		\item 
		outside the circle $C$ ($d_{C,p}>0$).
	\end{itemize}
	Since every circle $C$ in $\mathcal{C}_0'$ contains exactly 2 points $a,b$ of $P_0$ (and no points of $P_0$ in its interior),
	we have $d_{C,a}=d_{C,b}=0$ and $d_{C,p} > 0$ for every other point $p$ of $P_0$.
	Analogously, we define $d_{C,p}(\tau)$ at time~$\tau$.
	As $d_{C,p}(\tau)$ and $P(\tau)$ are both continuous functions,
	there exists $0 < \varepsilon_{C,p} \le \tau_{max}$
	such that $d_{C,p}(\tau)$ has the same sign for any $0 \le \tau \le \varepsilon_{C,p}$.
	We remark that $\varepsilon_{C,p}$ does not need to be maximal -- we just need some $\varepsilon_{C,p} > 0$ for our purposes.

\begin{figure}[htb]
		\centering
		\includegraphics[page=5]{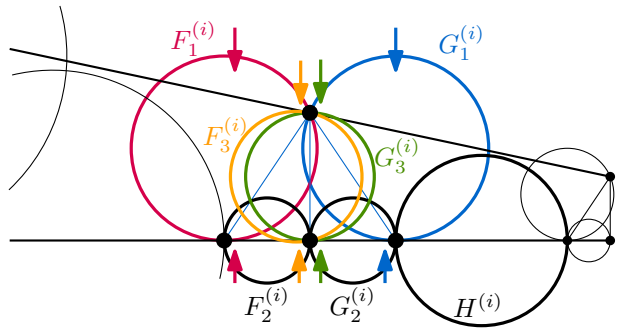}
		\caption{Analysis of a gadget and its corresponding circles. The colored arrows indicate the regions of the disks, which can be blocked by exterior points after the perturbation.}
		\label{fig:general_construction_all}
	\end{figure}

	Note that in the $i$-th gadget ($1<i<k$)
	the lower intersection point of the circles $F_1^{(i)}$ and $F_3^{(i)}$
	(as depicted in Figure~\ref{fig:general_construction_all})
	lies inside the convex hull of $P(\tau)$ at time $\tau=0$.
	Moreover, as this intersection point moves continuously on time,
	we can choose $\varepsilon_i > 0$ 
	such that at any time $0 \le \tau \le \varepsilon_i$
	this intersection point lies inside the convex hull.
	In an analogous manner, we can choose $\varepsilon_i' > 0$ for $1<i<k$
	such that at any time $0 \le \tau \le \varepsilon_i'$
	the lower intersection point of the circles $G_1^{(i)}$ and $G_3^{(i)}$
	(as depicted in Figure~\ref{fig:general_construction_all})
	lies inside the convex hull.
	
	Since we have a finite number of points and a finite number of circles,
	we can choose a common $ \varepsilon > 0$ small enough 
	such that  at any time $0 \le \tau \le \varepsilon$
	
	\goodbreak
	
	\begin{itemize}
		\item
		every circle in $\mathcal{C}(\tau)$ contains exactly 2 points of $P(\tau)$ (and no point in its interior), and
		\item
		no two exterior blocking areas overlap for $1<i<k$,
		except for the blocking areas of $F_1^{(i)}$ and $F_3^{(i)}$ on top, 
		and the blocking areas of $G_1^{(i)}$ and $G_3^{(i)}$ on top.
	\end{itemize}
	
	\goodbreak

	\subparagraph*{Analysis.}
	
	We first show that two points are required to block the circles $F_1^{(i)}$, $F_2^{(i)}$, and~$F_3^{(i)}$.
	If $F_1^{(i)}$ is blocked from above then we need at least a second
	point to block~$F_2^{(i)}$.
	Thus assume there is no point blocking $F_1^{(i)}$ from above.
	Since the above blocking area of $F_3^{(i)}$ is fully contained in $F_1^{(i)}$,
	the circle $F_3^{(i)}$ is also not blocked from above.
	Since the bottom blocking areas of $F_1^{(i)}$ and $F_3^{(i)}$ are disjoint,
	at least two blocking points have to be placed in $F_2^{(i)}$.
	As a consequence, two points are required to block $F_1^{(i)}$, $F_2^{(i)}$, and $F_3^{(i)}$.
	
	In an analogous manner one can show that two points are required to block the circles $G_1^{(i)}$, $G_2^{(i)}$, and $G_3^{(i)}$.
	It is easy to see, that
	\begin{itemize}
		\item
		the union of blocking areas of $F_1^{(i)}$, $F_2^{(i)}$, and $F_3^{(i)}$, 
		\item
		the union of blocking areas of $G_1^{(i)}$, $G_2^{(i)}$, and $G_3^{(i)}$, and
		\item
		the blocking area of $H^{(i)}$
	\end{itemize}
	are mutually disjoint.
	Consequently, at least five exterior blocking points are required for the $i$-th gadget ($1 < i < k$).
	Further, the blocking areas of the bottom circles of the first and last gadget ($F_2^{(1)}$, $G_2^{(1)}$, $H^{(1)}$, $F_2^{(k)}$, and $G_2^{(k)}$) are all disjoint from all other blocking areas.
	Hence, at least $5k-5$ points are required in total, 
	which completes the proof of Theorem~\ref{thm:genpos_thm}.

\section{Perturbation to general position}
\label{app:details}

Using linear algebra we give a formal proof for the existence of a sufficiently small $\varepsilon>0$ 
such that the point set $P(\tau)$ is in general position for $0 < \tau < \varepsilon$.
First, we show that for every triple of points $p,q,r$ from~$P$ 
there exists $\varepsilon_{pqr}>0$ such that
the perturbed points $p,q,r$ do not lie on a common line in $P(\tau)$ for $0 < \tau < \varepsilon_{pqr}$.
Second, we show that for every quadruple of points $p,q,r,s$ from~$P$ 
there exists $\varepsilon_{pqrs}>0$ such that
the perturbed points $p,q,r,s$ do not lie on a common circle in $P(\tau)$ for $0 < \tau < \varepsilon_{pqrs}$.
We can then find our desired $\varepsilon$ as the minimum among the finitely many
$\varepsilon_{pqr}$ and $\varepsilon_{pqrs}$ values.

The major idea in the following is that
collinearity and cocircularity can be expressed in terms of determinants 
(see e.g.\ Chapter~9 in \cite{CGbook}):
three points $p=(p_x,p_y)$, $q=(q_x,q_y)$, $r=(r_x,r_y)$ 
are collinear if and only if 
\[
\det 
\begin{pmatrix}
	1   & 1   & 1   \\
	p_x & q_x & r_x \\
	p_y & q_y & r_y \\
\end{pmatrix} 
= 0,
\]
and four points $p=(p_x,p_y)$, $q=(q_x,q_y)$, $r=(r_x,r_y)$, $s=(s_x,s_y)$ 
are cocircular if and only if 
\[
\det 
\begin{pmatrix}
	1   & 1   & 1   \\
	p_x & q_x & r_x \\
	p_y & q_y & r_y \\
	p_x^2+p_y^2 & q_x^2+q_y^2 & r_x^2+r_y^2 \\
\end{pmatrix} 
= 0.
\]

\subparagraph*{Collinearity:}
In~$P(0)$,
there are two lines which contain more than two points, namely the top and the bottom line. 
Any three points from one of these two lines are collinear, 
and moreover, any other triple of points is not collinear.
We have to cope with the perturbation 
which maps a point $p=(p_x,p_y)$
to $(x,y+\tau \sigma_p x^3)$
where $\sigma_p = +1$ (resp.\ $\sigma_p = -1$) 
if $p$ is a bottom point (resp.\ top point).
Hence, we define
\[
I_{pqr}(\tau) :=
\det 
\begin{pmatrix}
	1   & 1   & 1   \\
	p_x & q_x & r_x \\
	p_y+\tau \sigma_p p_x^3 & q_y+\tau \sigma_q q_x^3 & r_y+\tau \sigma_r r_x^3 \\
\end{pmatrix}
\]
for any three points $p,q,r$ of the construction.
Note that $I_{pqr}(\tau)$ is polynomial in $\tau$ and 
is thus either identically zero or has a finite number of roots.

Consider three bottom points $p,q,r$ (top points will be treated analogously).
We use the multilinearity of the determinant to write
\[
I_{pqr}(\tau)
=
\underbrace{
	\det
	\begin{pmatrix}
		1   & 1   & 1   \\
		p_x & q_x & r_x \\
		p_y & q_y & r_y \\
	\end{pmatrix} 
}_{=:A_{pqr}}
\,+\,
\tau \cdot 
\underbrace{
	\det
	\begin{pmatrix}
		1   & 1   & 1   \\
		p_x & q_x & r_x \\
		p_x^3 & q_x^3 & r_x^3 \\
	\end{pmatrix} 
}_{=:B_{pqr}}
=A_{pqr}+ \tau  \cdot B_{pqr}
\]
where the coefficients $A_{pqr}$ and $B_{pqr}$ are not depending on~$\tau$.
Therefore $I_{pqr}$ is either identically zero 
or has at most one root in $\tau = -\frac{A_{pqr}}{B_{pqr}}$ .

Next, observe that $B_{pqr}$ is the determinant of a generalized Vandermonde matrix, namely
\[
B_{pqr} = 
\begin{pmatrix}
	1   & 1   & 1   \\
	p_x & q_x & r_x \\
	p_x^3 & q_x^3 & r_x^3 \\
\end{pmatrix} 
= (q_x-p_x) \cdot (r_x-p_x) \cdot  (r_x-q_x) \cdot (p_x+q_x+r_x).
\]
Note that this identity can easily be verified using a computer algebra system such as SageMath~\cite{sage}.
Since all points have positive $x$-coordinates, the term $p_x+q_x+r_x$ is always positive.
Hence, the case $B_{pqr} = 0$ occurs if and only if 
two values of $p_x,q_x,r_x$ coincide. 
Since no two bottom points have the same $x$-coordinate,
we have $B_{pqr} \neq 0$.
Thus, $I_{pqr}(\tau)$ is not identically zero and 
we find a sufficiently small $\varepsilon_{pqr}>0$ 
such that the perturbed points $p,q,r$ are not collinear in $P(\tau)$ for $0<\tau<\varepsilon_{pqr}$.

\subparagraph*{Cocircularity:}
A similar argument can be used to deal with cocircularity.
For any four points $p,q,r,s$ of the construction, 
we define the polynomial $J_{pqrs}(\tau)$ as
\[
\det 
\begin{pmatrix}
	1   & 1   & 1 & 1   \\
	p_x & q_x & r_x & s_x \\
	p_y+ \tau \sigma_p  p_x^3 & 
	q_y+ \tau \sigma_q  q_x^3 & 
	r_y+ \tau \sigma_r  r_x^3 & 
	s_y+ \tau \sigma_s  s_x^3 \\
	p_x^2+(p_y+ \tau \sigma_p  p_x^3)^2 & 
	q_x^2+(q_y+ \tau \sigma_q  q_x^3)^2 &  
	r_x^2+(r_y+ \tau \sigma_r  r_x^3)^2 &  
	s_x^2+(s_y+ \tau \sigma_s  s_x^3)^2 \\
\end{pmatrix}.\\
\]

\noindent
To show that $J_{pqrs}(\tau)$ is not identically zero,
we write
\[
J_{pqrs}(\tau) = A_{pqrs} + \tau  \cdot B_{pqrs} + \tau^2  \cdot C_{pqrs} + \tau^3  \cdot D_{pqrs},
\] 
where $A_{pqrs}$, $B_{pqrs}$, $C_{pqrs}$, and $D_{pqrs}$ do not depent on $\tau$
and assume that all coefficients are zero,
i.e.,
the four points $p,q,r,s$ are cocircular in~$P(\tau)$ for every~$\tau$.
The constant term
\[
A_{pqrs} =
[\tau^0]\, J_{pqrs}(\tau) =
J_{pqrs}(0)
=
\det 
\begin{pmatrix}
	1   & 1   & 1 & 1   \\
	p_x & q_x & r_x & s_x \\
	p_y & q_y & r_y & s_y \\
	p_x^2+p_y^2 & q_x^2+q_y^2 & r_x^2+r_y^2 & s_x^2+s_y^2 \\
\end{pmatrix}
\]
is zero if and only if $p,q,r,s$ are cocircular in $P(0)$.
The coefficient of the cubic term is
\[
D_{pqrs} =
[\tau^3]\, J_{pqrs}(\tau) =
\det 
\begin{pmatrix}
	1   & 1   & 1 & 1   \\
	p_x & q_x & r_x & s_x \\
	\sigma_p p_x^2 & \sigma_q q_x^2 & \sigma_r r_x^2 & \sigma_s s_x^2 \\
	p_x^6 & q_x^6 & r_x^6 & s_x^6 \\
\end{pmatrix}.\\
\]
In the case $\sigma_p = \sigma_q = \sigma_r = \sigma_s = +1$
this
is again the determinant of a generalized Vandermonde matrix, which we can rewrite as
\[
D_{pqr} 
= (q_x-p_x) \cdot (r_x-p_x) \cdot  (r_x-q_x) \cdot (p_x+q_x+r_x) \cdot 
\left( 
\sum_{\substack{
		\alpha,\beta,\gamma,\delta \in \mathbb{N} \\ 
		\alpha+\beta+\gamma+\delta = 4}}
p_x^\alpha q_x^\beta  r_x^\gamma s_x^\delta
\right).
\]
Again, this identity can easily be verified using a computer algebra system.
Since all points have positive $x$-coordinates, 
the last term in the is always positive.
Hence, the case $D_{pqr} = 0$ occurs if and only if 
two values of $p_x,q_x,r_x,s_x$ coincide. 
The case $\sigma_p = \sigma_q = \sigma_r = \sigma_s = -1$ is analogous (with a negative sign).

Any four points in $P(0)$ from the top line (resp.\ the bottom line) are cocircular because
a line is a circle of infinite radius.
More precisely, the bottom points $\ell_1,m_1,r_1,\ldots,\ell_k,m_k,r_k$ lie on the $x$-axis 
and the top points~$t_i$ lie on the line  $\{(x,y) : 3x + 14y = 51\}$.
However, among each of such cocircular 4-tuple, 
all four points have distinct $x$-coordinates, i.e., $p_x,q_x,r_x,s_x$ are distinct and therefore $D_{pqrs}$ is non-zero in this case by the above analysis.

\begin{figure}[htb]
	\centering
	\includegraphics[page=1]{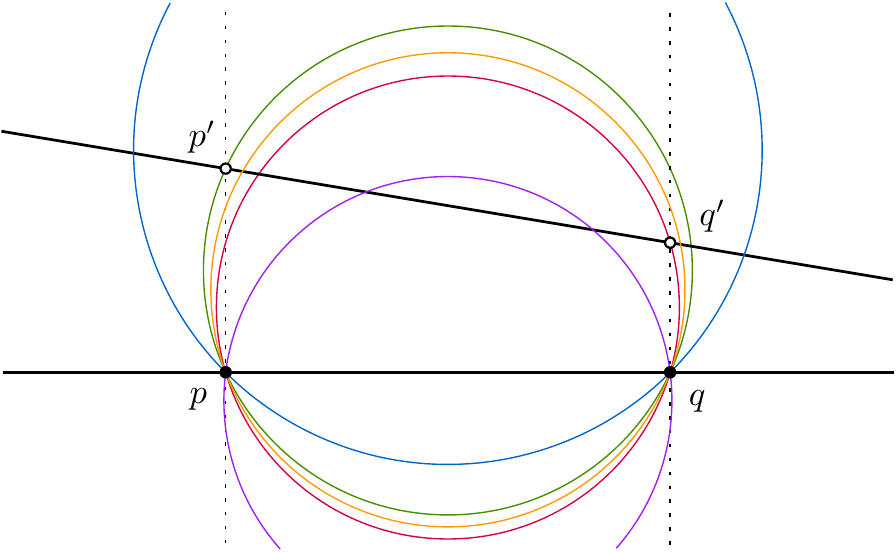}
	\caption{An illustration of cocircular points. Here $p'$ (resp.\ $q'$) denotes the point on the top line which lies directly above $p$ (resp.\ $q$).
	}
	\label{fig:cocircular}
\end{figure}

It remains to deal with the case, 
where the four points $p,q,r,s$ from~$P(0)$
lie on a common circle of finite radius.
Since any
three top points (resp.\ three bottom points) determine a line
and the fourth point would have to lie on this line,
it only remains to deal with the case of two bottom points (say $p,q$) 
and two top points (say $r,s$).
Moreover, we can relabel the four points 
so that $p$ is to the left of $q$ and $r$ is to the left of~$s$.
Because of the negative slope of the top line, 
there are five possibilities how the four points $p,q,r,s$ 
can occur from left to right, 
which are illustrated in Figure~\ref{fig:cocircular}: 

\goodbreak
\begin{itemize}
	\item $p_x < r_x < s_x < q_x$ (see the purple circle);
	\item $p_x < r_x < q_x = s_x$ (see the red circle);
	\item $p_x < r_x < q_x < s_x$ (see the yellow circle);
	\item $p_x = r_x < q_x < s_x$ (see the green circle);
	\item $r_x < p_x < q_x < s_x$ (see the blue circle).
\end{itemize}

In particular, we have $p_x \neq r_x$ or $q_x \neq s_x$.
We assume that $p_x,q_c,r_x$ are distinct; the case where $q_c,r_x,s_x$ are distinct will be treated in an analogous manner.
Since $p$ and $q$ are bottom points and $r$ is a top point,
they are not collinear and 
we have $I_{pqr}(0) \neq 0$. 
Since $p_x,q_c,r_x$ are distinct,
there exists $\tau^* \in \mathbb{R}$ 
such that the perturbed points $p,q,r$ are collinear in~$P(\tau^*)$.
Since the perturbed points $p,q,r,s$ are assumed to be cocircular at any time~$\tau$,
the four perturbed points $p,q,r,s$ are collinear in~$P(\tau^*)$.
This, however, is only possible if all four points have distinct $x$-coordinates,
that is,
$p_x,q_c,r_x,s_x$ are pairwise distinct.
Again, it follows that 
$D_{pqrs}$ is non-zero
and hence
$J_{pqrs}(\tau)$ is
not identically zero.
We again find $\varepsilon_{pqrs}>0$ 
such that the perturbed points $p,q,r,s$ are not cocircular in $P(\tau)$ for $0<\tau<\varepsilon_{pqrs}$.

\medskip 

This completes the proof that there is a sufficiently small $\varepsilon>0$
such that the point set $P(\tau)$ is in general position for $0 < \tau < \varepsilon$.

\section{Discussion and Further Related Work} 
\label{discussion}

The idea of blocking points can also be extended to other graph classes. For example,
Biedl et al.~\cite{BBIJKL2019}
investigated blocking sets of so-called $\Theta_6$-graphs, 
a structure related to Delaunay graphs:
In a $\Theta_6$-graph of a point set,
every pair of points shares an edge if
there is an empty equilateral triangle (instead of an empty disks).

From an algorithmic point of view, we can ask how fast a minimal blocking set can be computed.
For the general problem, where blocking points can also be placed in the interior of the convex hull of the Delaunay triangulation, this would help to identify cases where many blocking points are needed.
In fact, we tried several approaches to find a set of $n$ points which requires more than $n$ points to be blocked, but without success.
We therefore would not be surprised if Conjecture~\ref{conj:convex_n_blocking} always holds.
But even if Conjecture~\ref{conj:convex_n_blocking} is true, then there is still the algorithmic question how fast a blocking set of $n$ points can be found.

\begin{figure}[htb]
	\centering
	\includegraphics[page=1,width=0.95\textwidth]{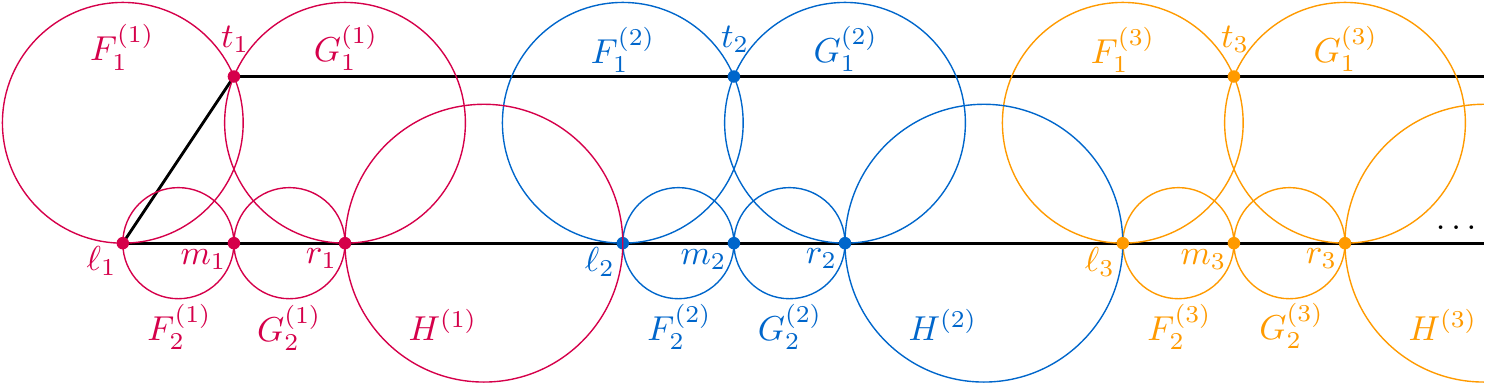}
	\caption{An alternative construction with isometric gadgets. 
		The red, blue, and yellow points and circles illustrate the first, second, and third gadget of the construction, respectively.
	}
	\label{fig:alternative_construction}
\end{figure}

The anonymous reviewers pointed out that the degenerate construction from Section~\ref{sec:collinear} can be improved as follows.
By removing the "middle" point $m_i$ from gadget~$i$ and replacing the circles $F_2^{(i)}$ and $G_2^{(i)}$ by
a circle $I_2^{(i)}$ with the segment $\ell_i r_i$ as diameter,
the constructed set of $3k$ points  (depicted in Figure~\ref{fig:alternative_construction}) 
requires $4k-2$ exterior-blocking points.
However, when making this construction non-degenerate via a perturbation as in Section~\ref{sec:genpos}, 
the number of required exterior-blocking points also drops significantly. 
\begin{theorem}
	\label{thm:collinear_thm}
	For $k \in \mathbb{N}$,
	there is a set $P$ of $3k$ points 
	that requires at least $4k-2$ exterior-blocking points.
\end{theorem}

\begin{figure}[htb]
	\centering
	\includegraphics[page=4,width=0.95\textwidth]{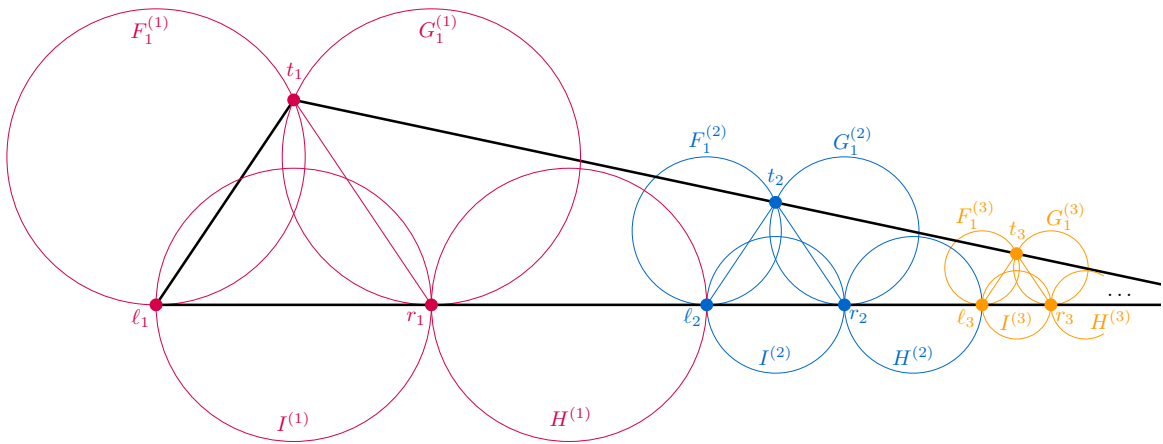}
	\caption{A degenerate construction with $3k$ points 
		where at least $4k-2$ exterior-blocking points are required. 
		The red, blue, and yellow points and circles illustrate the first, second, and third gadget of the construction, respectively.}
	\label{fig:construction_4_3}
\end{figure}

A reviewer also pointed out that the gadgets in the degenerate construction need not to be scaled.
Figure~\ref{fig:alternative_construction} gives an illustration
of the alternative construction. 
However, when making this construction non-degenerate via a perturbation as in Section~\ref{sec:genpos}, the number of required exterior-blocking points significantly drops because, for $1<i \le k$, the circles $G_{3}^{(i)}$ can be blocked by points that are to the left of~$t_i$ and slightly above~$t_{i-1}t_i$.
Also note that, in contrast to our construction from Section~\ref{sec:collinear}, here the four points $m_i,t_i,m_j,t_j$ lie on a common circle for every $1 \le i<j \le k$.

{
	\small
	\bibliography{references}
	\bibliographystyle{abbrv-url-doi}
}

\end{document}